\documentclass{an_art}
\usepackage{psfig}
\usepackage{graphicx}
\begin{document}
\begin{titlepage}
\setcounter{page}{1}
\headnote{Astron.~Nachr. ~00 (0000) 0, 000--000}
\makeheadline
\title{Stellar Population Analysis from Broad-Band Colours}
\author{
{\sc Hamed \ Abdel-Hamid}, Cairo, Egypt. \\
\medskip
{\small National Research Institute of Astronomy and Geophysics} \\
\bigskip
{\sc Peter Notni}, Potsdam, Germany.\\
\medskip
{\small Astrophysikalisches Institut Potsdam}\\
}

\date{Received .....; accepted .....}
\maketitle

\summary
We have developed an analytical method to investigate the stellar populations 
in a galaxy using the broad-band colours. The method enables us to determine 
the relative contribution, spatial distribution and age for different stellar 
populations and gives a hint about the dust distribution in a  galaxy. \\
We apply this method to the irregular galaxy NGC 3077, using CCD images in U, B , V and R filters.
END
\keyw
Galaxies: Stellar populations -- Galaxies: individual (NGC 3077) 
END
\end{titlepage}
\section{Introduction}
  The investigation of stellar populations in  a galaxy using broad-band 
colours is based on an analysis 
of the distribution of stars in the colour magnitude diagram (CMD), and  the 
comparison of the observed CMD with a set of model CMDs.
 In more distant galaxies, individual stars are  difficult to observe, 
and hence one must interpret the {\em integrated\/} light in terms of the 
stellar populations and use colours of model populations for comparison.  
Papers using the last-mentioned approach 
are becoming more frequent recently. For instance, Kong et al.
(2000) used a set of 13 intermediate-band colours to get information on the 
distribution of age, extinction and metallicity in M81, assuming a  
population of stars of equal age in each pixel.     
Bell et al (1999) and 
Abraham et al (1999) analysed the pixel-by-pixel colour distribution 
to get the overall star formation history, 
comparing the colours with theoretical predictions for populations with 
various time-scales $\tau$ of the star forming process.

In contrast, Notni and Bronkalla (1984) derived the population
distribution in the inner regions of M82 assumimg {\em two distinct}\/ 
populations  (each with small $\tau$) in each pixel in their analysis of 
photographic data. Similarly, 
the integrated colour for a sample of dwarf galaxies was interpreted 
by Almoznino and Brosch (1998) 
as a combination of a young (3-13Myr) and an old (1-2 Gyr) stellar populations. 
They get their conclusion by calculating some combinations for the integrated 
light of two populations with different ratio. 
 We start,  the other way round, from the observed light in every point in the 
galaxy and try to decompose  this light into its contributions from the two 
stellar populations by solving the equations which describe the combined 
colours.   
The method enables us  to estimate the age and the extinction of the population
 I and the contribution of both populations I and II to  the observed light.

In the present paper, the mathematical  presentation of the method is given in 
section 2. A short error analysis in the determination of age and relative 
contribution of different populations is discussed in section 3. 
In section 4, an application of the method using CCD observations of the 
irregular galaxy NGC 3077 is presented. 
The conclusion will be given in section 5.
The observations used in this work and their reduction are presented in  
our preceeding paper in this issue (Abdel-Hamid \& Notni 2000, HN1 hereafter).

\section{Derivation of the relative abundance of two populations}
\begin{figure}[htbp]
\begin{minipage}[b]{110mm}
\psfig{figure=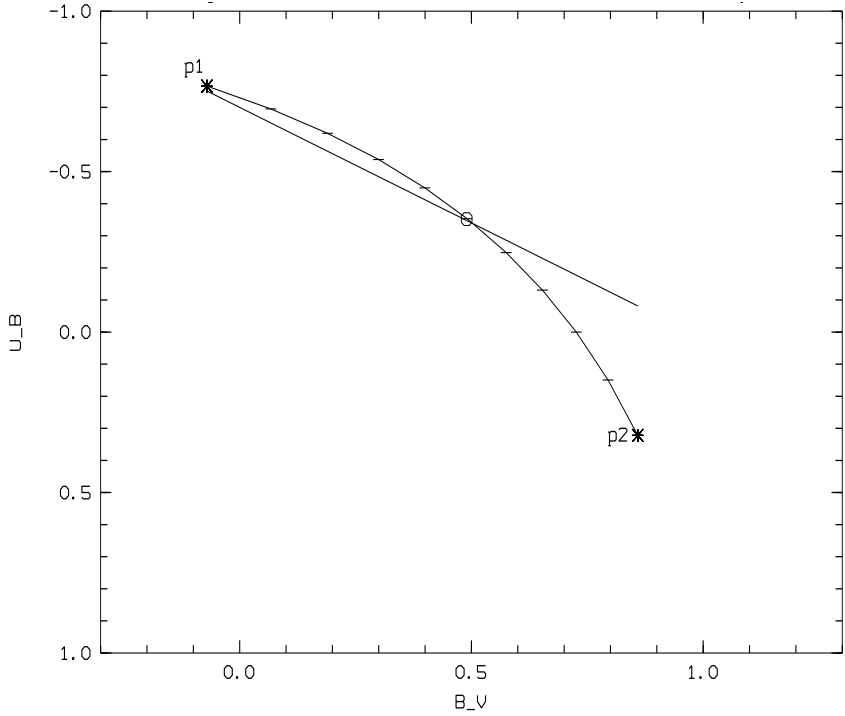,bbllx=49pt,bblly=60pt,bburx=560pt,bbury=790pt,height=95mm,width=115mm,angle=-90,clip=}
\end{minipage}
\hfill
\parbox[b]{55mm}{
\caption{ A mixing track in the UBV colour-diagram.  Population I (p1)
 and population II (p2) mix and  produce the observed colour along the 
curved track. 
 Ticks on the mixing track represent different contributions of the populations
in the observed light ($\beta$), step = 0.1. The open circle marks an observed colour, in which 50\% of its total
 light comes from the population I (p1). The solid line is a reddening vector.}
\label{mix}}
\end{figure}

Let us assume that the observed light of a galaxy comes from only two stellar 
populations with different colours, i.e. with different age, metallicity, etc. 
A graphical presentation of the line along which the 
colour changes if two given populations mix in various proportions is given in 
Figure \ref{mix}.  P2 refers to the colour of an old population and P1 to the colour of the young population.
Both P1 and P2 mix along the line which connects them to give the observed colour.

In order to derive the required relations which describe the mixing-tracks of the two stellar populations, we start with some definitions.
The suffixes m, I and II refer to total (measured), population I and population II intensity, respectively. The colour indices are defined with respect to that of population II.
\\
\begin{tabbing}
 $\triangle(U-B)_m$ \= = \= $(U-B)_m - (U-B)_{II}$, \quad\quad \= $\triangle(U-B)_I$ \= =  \= $(U-B)_I - (U-B)_{II}$\\
 $\triangle(B-\hspace{-.3mm}V)_m$ \= = \= $(B-V)_m - (B-V)_{II}$, \quad\quad \=$\triangle(B-V)_I$ \= =  \= $(B-V)_I - (B-V)_{II}$\\
 $\triangle(B-R)_m$ \= = \= $(B-R)_m - (B-R)_{II}$, \quad\quad \= $\triangle(B-R)_I$ \= =  \= $(B-V)_I - (B-V)_{II}$\\
 $\triangle(B-\hspace{1mm}I)_m$ \= = \= $(B-\hspace{1mm}I)_m - (B-\hspace{1mm}I)_{II}$, \quad\quad \= $\triangle(B-\hspace{1mm}I)_I$ \= = \= $(B-\hspace{1mm}I)_I - (B-\hspace{1mm}I)_{II}$\\
\end{tabbing}
Then we define the following intensity relations:
\\
\[ \frac{u_m}{b_m}  =  \frac{u_I \cdot e^{-\tau_u} +  u_{II}}{b_I\cdot e^{-\tau_b} + b_{II}} \quad\mbox{,}\quad
\frac{b_m}{v_m}  =  \frac{b_I \cdot e^{-\tau_b} +  b_{II}}{v_I\cdot e^{-\tau_v}+  v_{II}}  \]
\[ \frac{b_m}{r_m}  =  \frac{b_I \cdot e^{-\tau_b} + b_{II}}{r_I\cdot e^{-\tau_r} +  r_{II}} \quad\mbox{,}\quad
 \frac{b_m}{i_m}  =  \frac{b_I \cdot e^{-\tau_b} +  b_{II}}{i_I\cdot e^{-\tau_i}+  i_{II}} \]
where u,b,v,r and i are the {\em dereddened\/} intensities in the corresponding
 band and for the corresponding population, while that with suffix $m$ are the observed total light intensities.  $\tau_b$ is the optical thickness in the B band.\\
Note that only population I is assumed to suffer extinction, equivalent to the assumption of a typical galaxy structure  governed by a flat young population associated with dust structures, embedded in an old population mainly outside the extinction layer.
\\
Converting the above equations to magnitudes we get
 \[ (U-B)_m = (U-B)_{II} - 2.5\log\left( \frac{ \alpha X_0
e^{-\tau_u} + 1}{ \alpha e^{-\tau_b} + 1}\right) \]
 \[ (B-V)_m = (B-V)_{II} - 2.5\log\left( \frac{ \alpha  e^{-\tau_b} +
1 }{ \alpha X_1 e^{-\tau_v} + 1 } \right) \]
 \[ (B-R)_m = (B-R)_{II} - 2.5\log\left( \frac{ \alpha  e^{-\tau_b} +
  1 }{ \alpha X_2 e^{-\tau_r} + 1 } \right) \]
 \[ (B-I)_m = (B-I)_{II} - 2.5\log\left( \frac{ \alpha  e^{-\tau_b} +
1 }{ \alpha X_3 e^{-\tau_i} + 1 }\right) \]
 with
 \[ \alpha = \frac{\beta_b}{e^{-\tau_b} - \beta_b \cdot e^{-\tau_b}} \quad\mbox{
 and}\quad \beta_b =\frac{b_I\cdot e^{-\tau_b}}{b_m} = \frac{b_I\cdot e^{-\tau_b
}}{b_I\cdot e^{-\tau_b} + b_{II}}\]
\[ \frac{u_I}{b_I} = X_0 \cdot \frac{u_{II}}{b_{II}}
\quad\mbox{,}\quad \frac{b_{II}}{v_{II}} = X_1 \cdot \frac{b_{I}}{v_{I}}
\quad\mbox{,}\quad \frac{b_{II}}{r_{II}} = X_2 \cdot \frac{b_{I}}{r_{I}}
\quad\mbox{and}\quad \frac{b_{II}}{i_{II}} = X_3 \cdot
\frac{b_{I}}{i_{I}}\]
and
 \[ X_0 = dex(-0.4\triangle(U-B)_I) \quad\mbox{,}\quad X_1 =
dex(0.4\triangle(B-V)_I)\]
\[  X_2 = dex(0.4\triangle(B-R)_I) \quad\mbox{,}\qquad X_3 =
dex(0.4\triangle(B-I)_I)\]
Here  $e^{-\tau_{\lambda}} = dex(-0.4\cdot A_{\lambda})$, and using a total to 
 selective extinction ratio of $ R_v = 3.1$, the colour excess relations for other
 wavelength bands is
$E_{u-b} = 0.72E_{b-v}$, $E_{v-i} = 1.25E_{b-v}$ and $E_{v-r} = 0.62E_{b-v}$ 
 (Grebel \& Roberts 1995).\\
 Note that the X's are functions of the colours of both populations I and II.\\
 We get:
\begin{eqnarray}
  (U-B)_m  &=& (U-B)_{II} -2.5\log\left( 1 - \beta_b + \beta_b \cdot 10^{(-0.288
 E_{b-v})} \cdot 10^{(-0.4\triangle(U-B)_I)}\right)\nonumber \\
 (B-V)_m  &=& (B-V)_{II} + 2.5\log\left( 1 - \beta_b + \beta_b \cdot 10^{(0.4 E_
{b-v})} \cdot 10^{(0.4\triangle(B-V)_I)}\right)\nonumber \\
 (B-R)_m  &=& (B-R)_{II} + 2.5\log\left( 1 - \beta_b + \beta_b \cdot 10^{(0.648
E_{b-v})} \cdot 10^{(0.4\triangle(B-R)_I)}\right)\nonumber \\
 (B-I)_m &=& (B-I)_{II} + 2.5\log\left( 1 - \beta_b + \beta_b \cdot 10^{(0.9 E_{
b-v})} \cdot 10^{(0.4\triangle(B-I)_I)}\right) \nonumber \\
\nonumber
\end{eqnarray}

A common wavelength band (e.g. B in our case) in the colour indices is essential
 to get the relative abundance of the populations in the observed intensity. \\
In general form the observed colour index is given as:\\
\begin{equation}
\label{gen}
(B-\lambda)_{m} = (B-\lambda)_{II} + 2.5\log\left( 1 - \beta_b + \beta_b \cdot 1
0^{c\cdot E_{b-v)}} \cdot  10^{(0.4\triangle(B-\lambda)_I)}\right)
\end{equation}
where the colour indices are a difference between a certain wavelengthband 
 ($\lambda$) and the B-band. $c$ is a constant, its value depends on the colour excess
 relations for the wavelength bands.\\
From the above equations, the required relation for the relative light-contribution of population I, $\beta_b$, is:
\begin{eqnarray}
\label{betaeq}
\beta_b & = & \frac{ 1 - A}{ 1 - X_0 \cdot dex(-0.288 E_{b-v})} \nonumber\\
& = & \frac{ 1 - B}{ 1 - X_1 \cdot dex(0.4 E_{b-v})}  \\
& = & \frac{ 1 - C}{ 1 - X_2 \cdot dex(0.648 E_{b-v})} \nonumber \\
& = & \frac{1 -  D}{1 - X_3 \cdot dex(0.9 E_{b-v})} \nonumber
\end{eqnarray}
with
\[ A  = dex(-0.4\triangle (U-B)_m) \qquad\mbox{,}\qquad  B  =  dex(0.4
\triangle (B-V)_m)  \]
\[ C  =  dex(0.4\triangle (B-R)_m) \qquad\mbox{,}\qquad  D  =  dex(0.4
\triangle (B-I)_m) \]
 A, B, C and D are functions of the observed and population II colours.

If we replace in equations \ref{gen} the colour indices of both populations I and II by the age of each of them, then these equations contain only four
 unknowns (age of population I, age of population II, $\beta$ and $E_{b-v}$). 
 The system becomes mathematically solvable, whenever four observed colour indices are used.
To provide the age dependence of the broad-band colours of the stellar
 populations, a  population synthesis model is  used.
Having only three colour indices, the system of equations cannot be solved
 without further assumptions, however. In many cases, including NGC3077, 
the colour of population II, i.e. its age, can be determined independently 
from some outer regions of the galaxy. This reduces the number of unknowns 
to 3 and the system becomes solvable already with only three measured colours.  
In NGC3077, the colour of population II is assumed to be the mean of the 
data clustering around U-B = 0.3 mag, B-V = 0.85 mag and B-R = 1.4 mag 
in the colour diagrams, corresponding to an age of 2..4x10$^9$ years, 
see  Figure \ref{syP} and HN1.  
An even more restricted solution of the problem was used by Notni and Bronkalla 
(1984) to analyse the distribution and extinction of two populations of 
known reddening-free index [Q=(U-B)-0.65(B-V)] in the galaxy  M82 from only 
two colour indices. 
\section{Short error analysis}
\begin{figure}[htbp]
\setlength{\unitlength}{1.0cm}
\begin{minipage}[t]{8.0cm}
\begin{picture}(8.5,8.5)
\psfig{figure=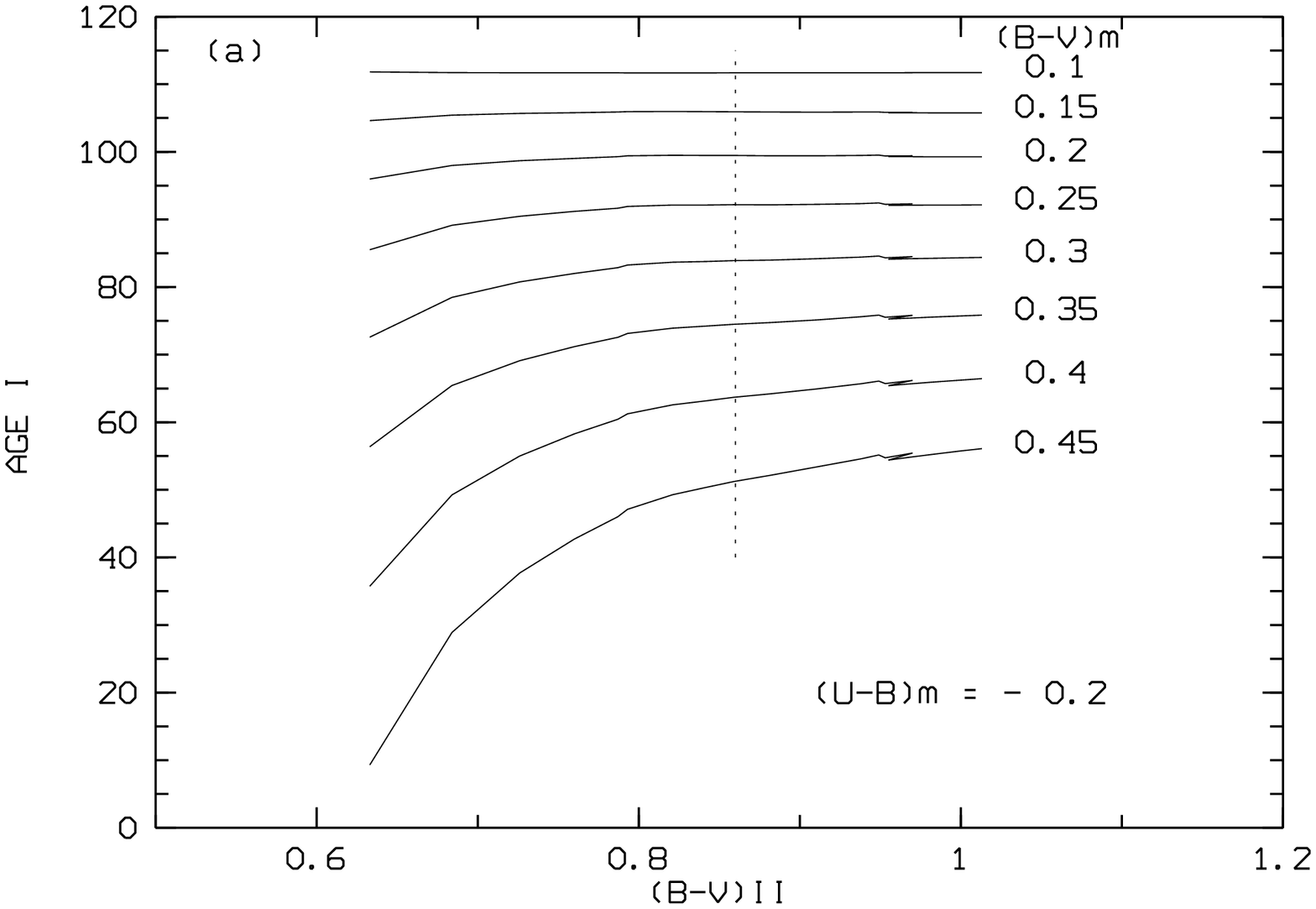,bbllx=108pt,bblly=69pt,bburx=515pt,bbury=692pt,height=80mm,width=80mm,angle=-90,clip=}
\end{picture}
\end{minipage} \hspace{1cm}
\begin{minipage}[t]{8.0cm}
\begin{picture}(8,8)
\psfig{figure=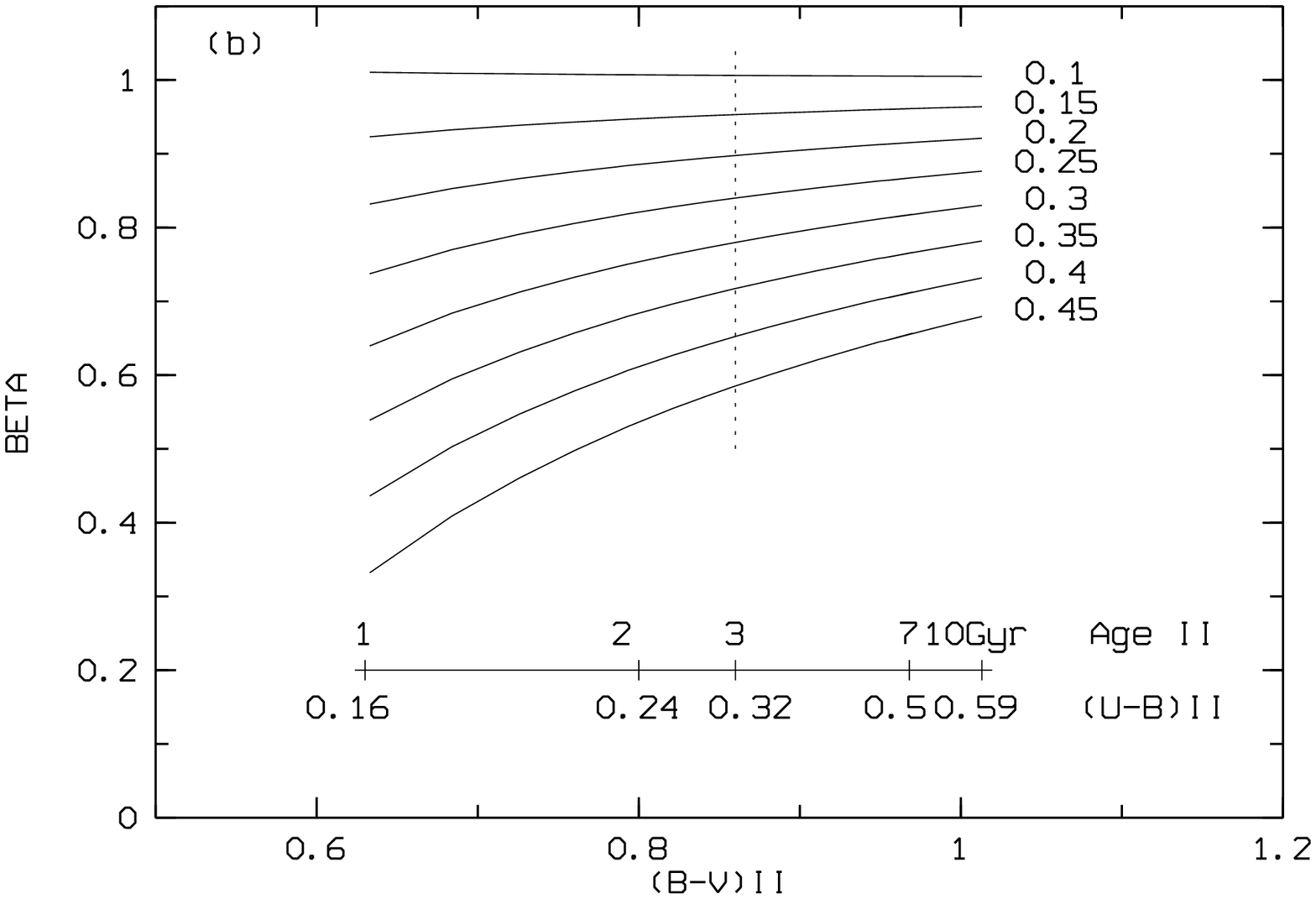,bbllx=108pt,bblly=69pt,bburx=515pt,bbury=692pt,height=80mm,width=80mm,angle=-90,clip=}
\end{picture}
\end{minipage}
\caption{\label{errbv} The effects of change of the assumption of the colour 
(or, equivalently, age) of population II  for eight test data 
points with constant (U-B)m = -0.2
 and (B-V)m ranging from 0.1 to 0.45, on (a) the estimation of the age
 of population I, in Myr, and (b) its contribution percent $\beta$, in the observed
intensity. The population II is represented, along the abscissa, by its age or,
alternativly, its colours, assumed changing along the age-line 
for z=0.02 of the BCF94 model.
 The dotted line gives the influence of errors in the {\em observed\/} 
mixed colours on the
 deduced parameters $age(popI)$ and $\beta$, see  text.}
\end{figure}

\begin{figure}[htbp]
\setlength{\unitlength}{1.0cm}
\begin{minipage}[t]{8.0cm}
\begin{picture}(8,8)
\psfig{figure=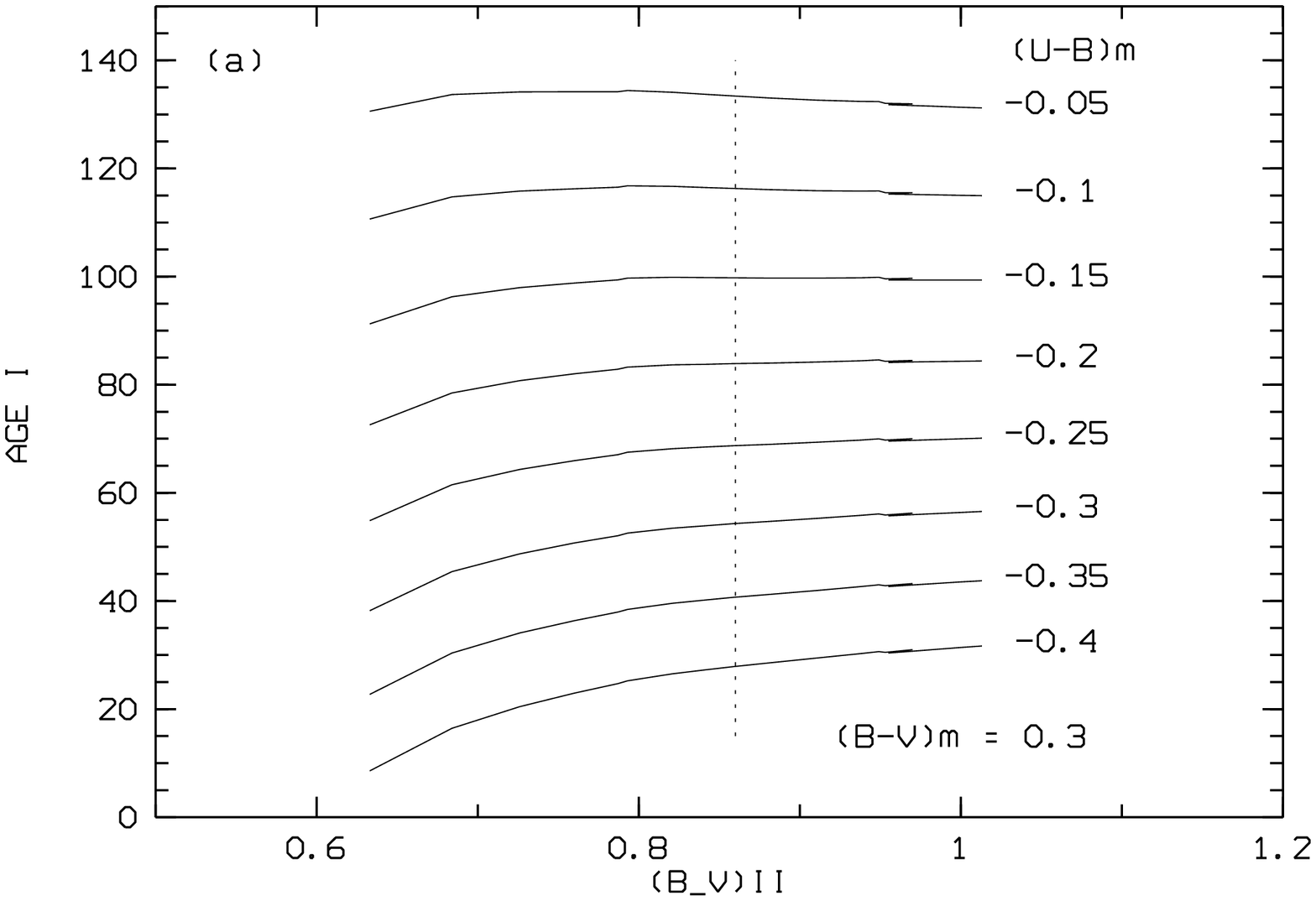,bbllx=108pt,bblly=69pt,bburx=515pt,bbury=692pt,height=80mm,width=80mm,angle=-90,clip=}
\end{picture}
\end{minipage} \hspace{1cm}
\begin{minipage}[t]{8.0cm}
\begin{picture}(8,8)
\psfig{figure=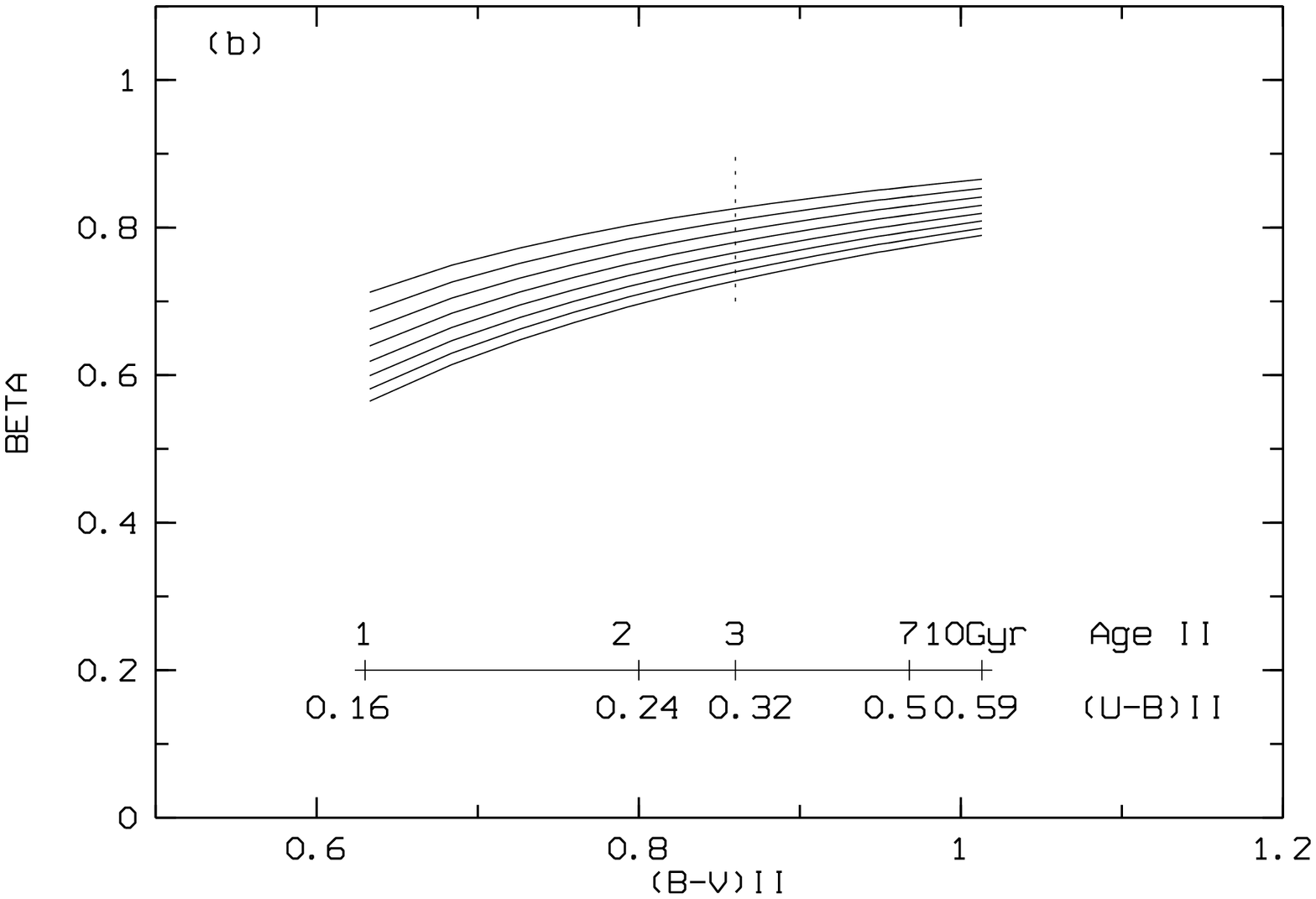,bbllx=108pt,bblly=69pt,bburx=515pt,bbury=692pt,height=80mm,width=80mm,angle=-90,clip=}
\end{picture}
\end{minipage}
\caption{\label{errub} As Figure \ref{errbv} , but for a set of observed points
 with constant (B-V)m = 0.3 and (U-B)m ranging from - 0.05 to - 0.40.}
\end{figure}

For the typical assumption of a known colour (or age) of population II, 
we check the influence
 of errors in this assumption on the deduced parameters $age(popI)$ and $\beta$,
assuming for this test zero extinction. We use the UBV diagram and the 
theoretical age line of Bressan et al. 1994 (hereafter BCF94), illustrated in 
Figure \ref{mt2}.
We try to estimate the error in the estimation of the age of the population I, 
 (the left panels in the Figures \ref{errbv} and \ref{errub}), and on  the ratio
 $\beta$ (the right panels).

If the assumed age of population II goes  from 2 to 10 Gyr, 
 (corresponding to a range in $(B-V)_{II}$ from 0.80 to 1.01 or $(U-B)_{II}$ from 0.24 to 0.59),
 this will produce an error in the estimation of the age of population I
 between 0 (uppermost curves in Figures \ref{errbv} and \ref{errub}) and 9 Myr 
 (lowest curves), and  an error in the ratio $\beta$ between 0 and 14\% 
respectively, where
 the amount of error depends on the location of the test points in the 
two-colour diagrams. 
At lower ages, if the  estimated population II age moves  
  between 1 and 2 Gyr, the results are less certain; 
the error may be as high as 
a factor of four in the age of population I and 20\% for $\beta$. 

In the region of practical interest, where the age of population II is 3$\pm 1$ Gyr,
 the estimation of the age of population I is accurate to about 4 Myr and 
 $\beta$ by about 7\% in the average. .

The same Figures can also be used to get the influence of errors in the 
 {\it observed mixed} colours on these parameters.  The dotted line in Figures \ref{errbv}
 and \ref{errub} is used for this; it is valid for a combination of an old population
 with an age of 3 Gyr and a young population 
(characterized by age or  its contribution $\beta$), 
which will give the observed colours $(B-V)_m$ and $(U-B)_m$.
 The movement of the observed colours by 0.1 on the dotted line, i.e. if there is
 an error of 0.1 in the observed colours, leads to an error of the estimation of
 $\beta$ by an amount of 10\% and of the age of population I by 20 Myr in the average.

\section{Applying the method to the observations}
\begin{figure}[htbp]
\setlength{\unitlength}{1.0cm}
\begin{minipage}[htbp]{8.4cm}
\begin{picture}(8.4,8.4)
\put(4.2,-.4){{\bf B-V}}
\psfig{figure=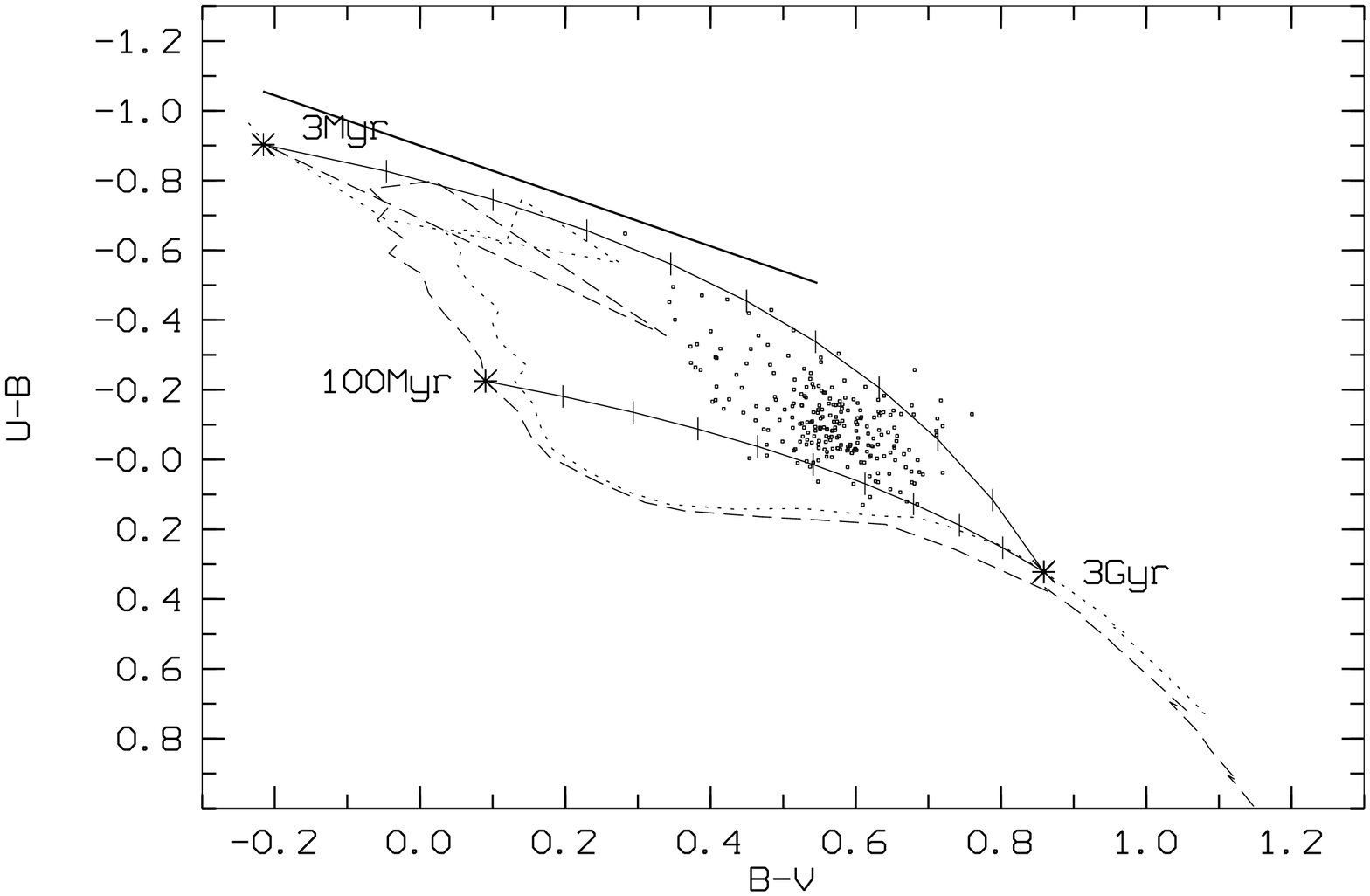,bbllx=79pt,bblly=57pt,bburx=502pt,bbury=732pt,height=80mm,width=83mm,angle=-90,clip=}
\end{picture}
\end{minipage} \hspace{1cm}
\begin{minipage}[p]{8.4cm}
\begin{picture}(8.4,8.4)
\put(4.2,-.4){{\bf B-R}}
\psfig{figure=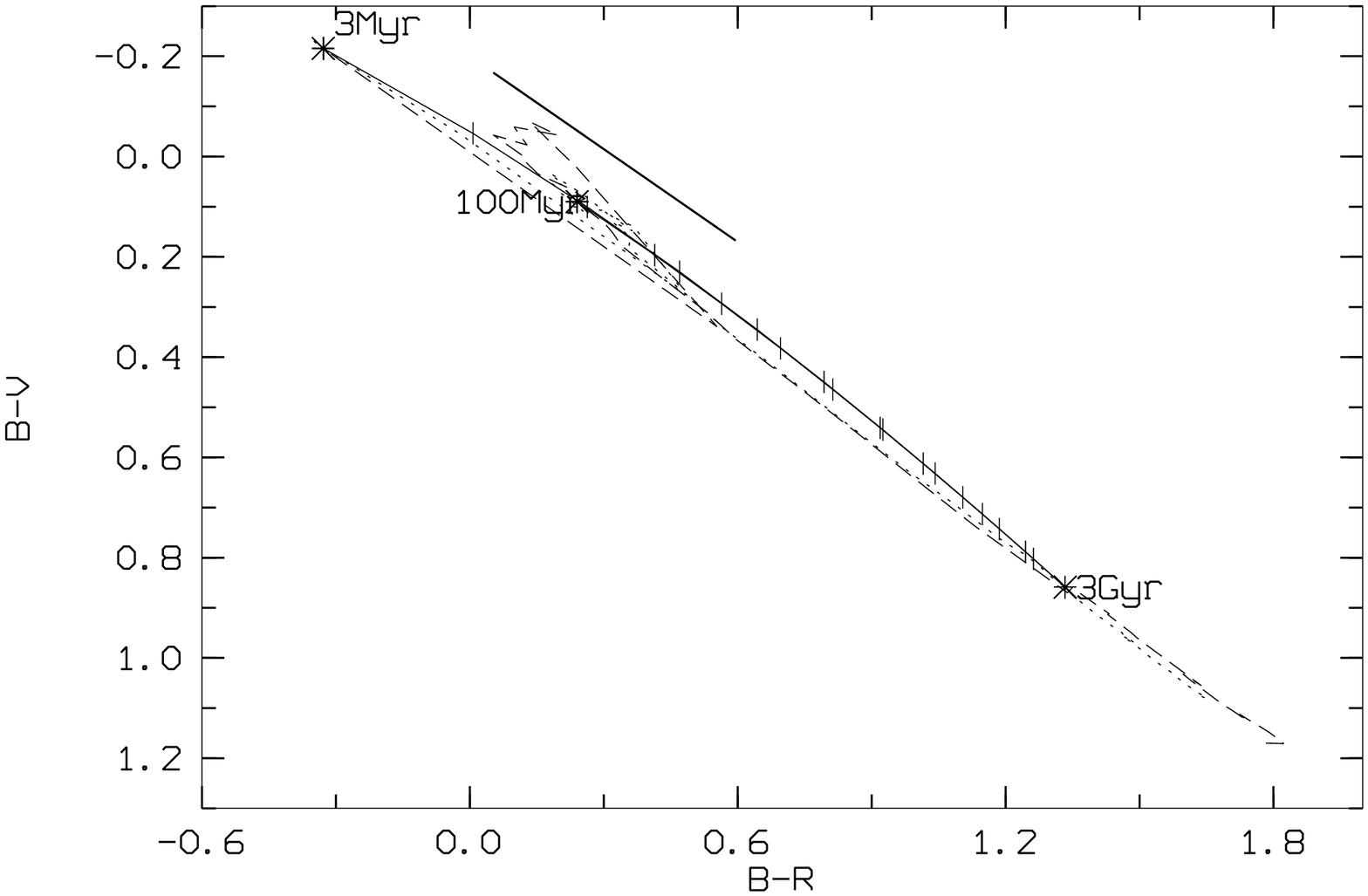,bbllx=79pt,bblly=57pt,bburx=502pt,bbury=732pt,height=80mm,width=83mm,angle=-90,clip=}
\end{picture}
\end{minipage}
\vspace{.1mm}
\caption{\label{mt2} The mixing-tracks in two different colour diagrams, 
with two mixing curves joining an old population (3Gyr) with two age limits 
of the young population (3, 100 Myr). The points in the left  panel are 
observed colours in the central region. Ticks are $\beta$ values, step 0.1 (see text).  Dotted and long dashed curves are population synthesis models with Z = 0.02 and Z = 0.05, 
respectively (BCF94).  The solid thick line is a reddening vector.}
\end{figure}

We analyse  the observational data for NGC 3077  using as an age reference 
line the models of BCF94, as illustrated in Figure \ref{mt2}. 

In the UBV-diagram, the mixing curve of the 
two populations is inclined to the age line, 
but about parallel to the reddening line.
This  leads to an easy estimate of the age of population I, 
but it is  impossible to distinguish reddening from
 the effect of the varying contributions of the two populations 
($\beta$, ticks in the Figures)
 using only these colour indices (U-B and B-V). For every age
 estimation there are many possible combinations of $\beta$ and $E_{b-v}$.
 Therefore, another two-colour diagram is required in order to get discrete values
  of age, $\beta$ and $E_{b-v}$ for each observed colour. We use the 
two-colour diagram B-V / B-R for this purpose.
 
A MIDAS iteration subroutine is used to solve the system of equations \ref{gen} 
 as follows: for each age value  in some age range of the young population I, 
its colour indices are read as input parameters; for population II   constant 
colours corresponding to an age of 3x$10^9$ yrs (z=0.02) are used. 
 Then we calculate three $\beta$ images, one for each observed colour,  and 
the standard deviation of these values in every pixel, the ($\sigma_{\beta}$) 
image.
  This is done for 14 different age values of population I, ranging 
from 4x$10^6$ to $10^8$ yrs.
From these 14  $\sigma_{\beta}$ images the subroutine looks for the 
minimum value of $\sigma_{\beta}$ at each pixel and writes its corresponding 
age, ${\beta}$ and $\sigma_{\beta}$, at this position in three new images, 
named for example $Age$0, $\beta$0 and $\sigma_{\beta}$0 (for an assumed 
value of extinction $E_{b-v}$ = 0).
 In a second step the program repeats these calculations for a range of extinction values, in terms of $E_{b-v}$ from 0 to 1.0, step 0.1.
 This will give for each  parameter set ($age$, ${\beta}$ and $\sigma_{\beta}$) 
10 images. From these 10 $\sigma_{\beta}$ images the program searches again 
the minimum $\sigma_{\beta}$ and writes this value  and the 
corresponding $age$, $\beta$ and $E_{b-v}$ values in the corresponding 
position in four new images.
The three images  $\beta$, $age$ and $E_{b-v}$ are the required solution of 
equation \ref{gen} for each pixel.
 The $\beta$ image gives at every pixel the contribution
 of population I in the observed light (in B filter in this case), while the
$age$ and $E_{b-v}$ images give the distribution of the age of population I and
 its associated extinction in the field of interest.

\begin{figure}
\setlength{\unitlength}{1cm}
\begin{minipage}[htbp]{8.4cm}
\begin{picture}(8.4,8.4)
\put(4.0,-.4){{\bf B-V}}
\psfig{figure=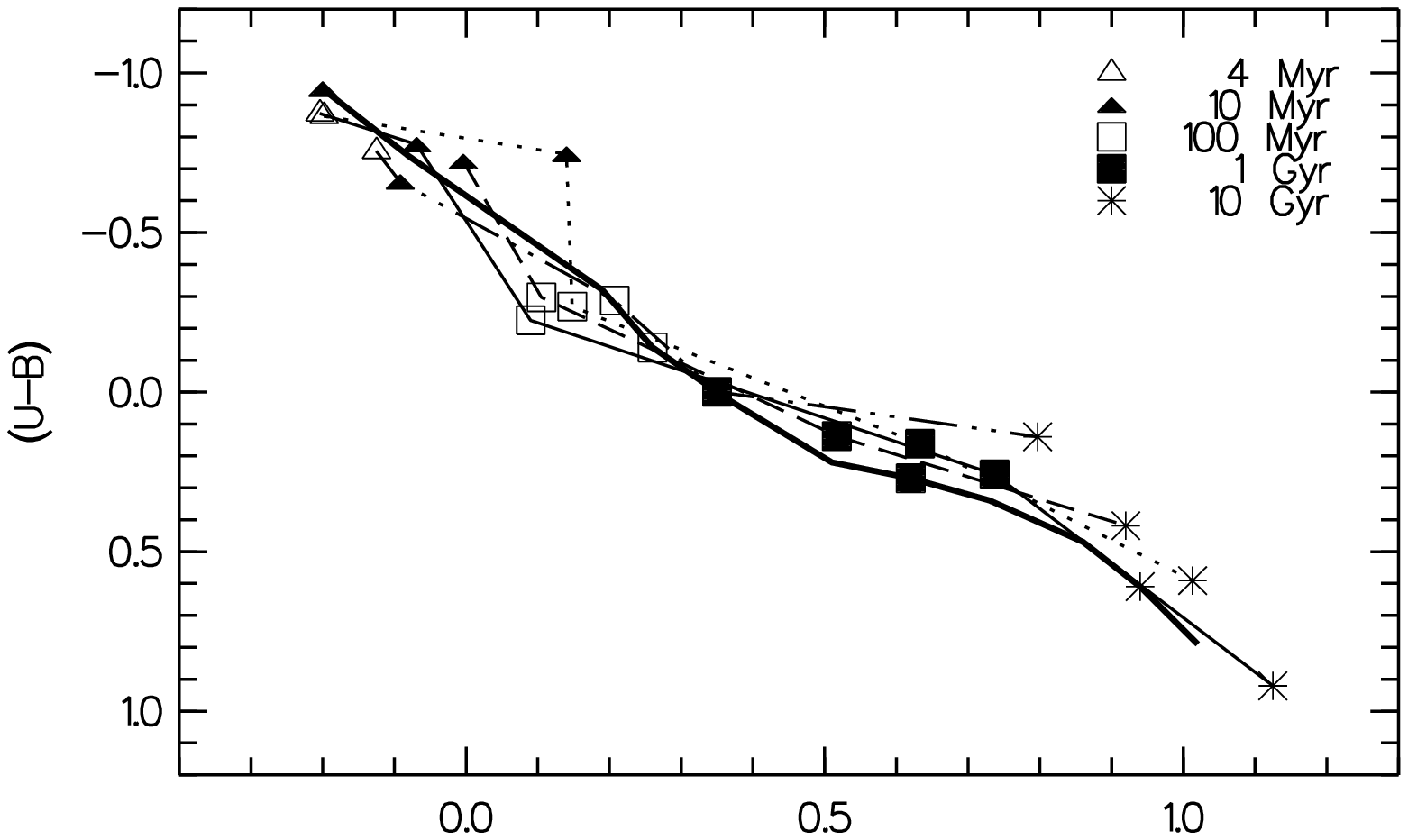,bbllx=180pt,bblly=135pt,bburx=459pt,bbury=585pt,height=80mm,width=83mm,angle=-90,clip=}
\end{picture}\par
\vspace{0.1cm}
\caption{\label{syP}Two-colour diagram of the population synthesis models by 
BCF94 and LT78 in comparison. The symbols refer to the age of the populations, 
different lines are different  metallicities as follows: solid (Z = 0.05), dotted (Z = 0.02), short dashed (Z = 0.008) and long dash-doted (Z = 0.001) for BCF94. The thick solid line is the Larson \& Tinsley model 1978 with solar metallicity, Z = 0.02 (for comparison).}
\end{minipage}\hfill
\begin{minipage}[htbp]{8.4cm}
\begin{picture}(8.4,7.4)
\put(3.0,-0.4){\bf Semi-major axis a ($\prime\prime$)}
\psfig{figure=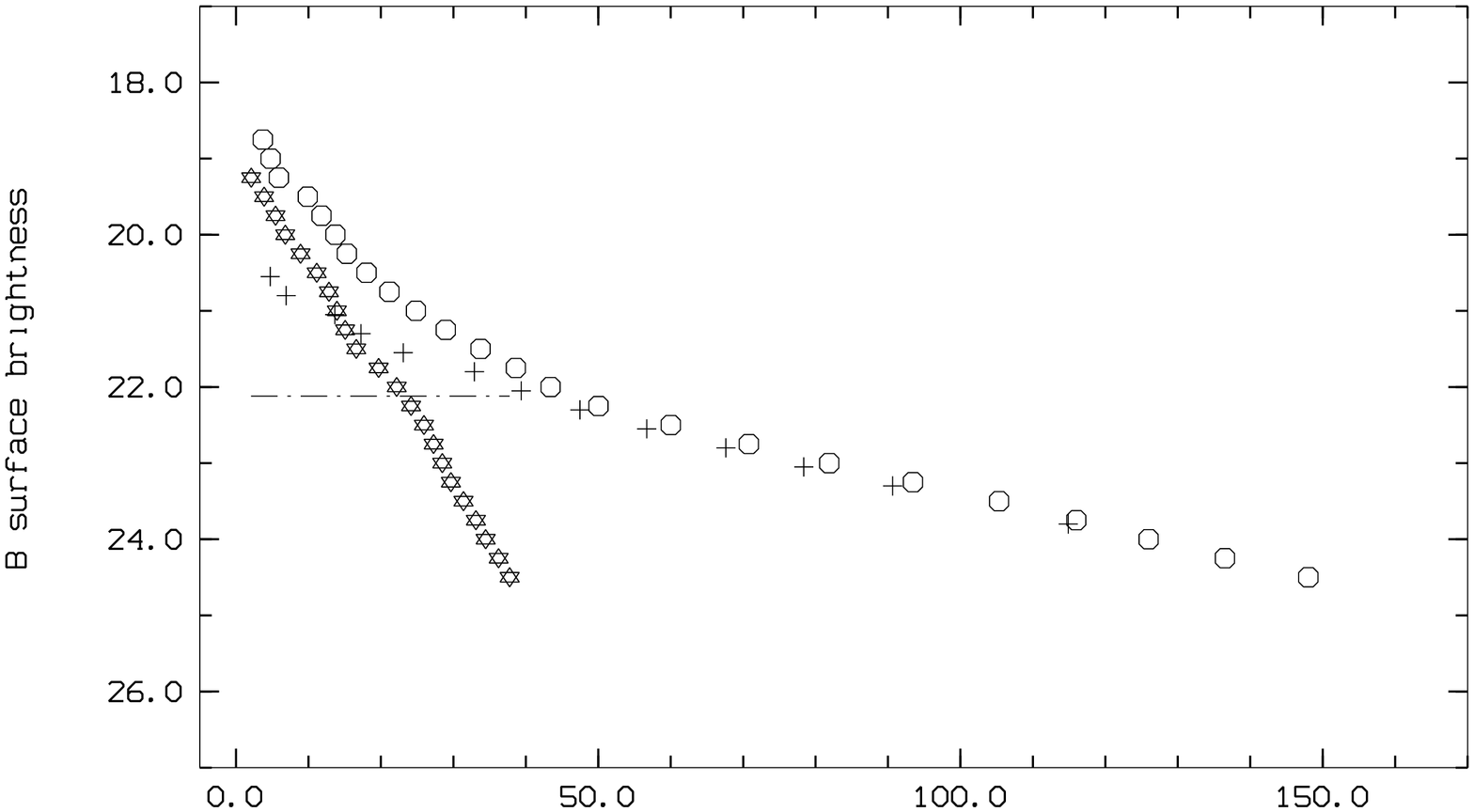,bbllx=97pt,bblly=63pt,bburx=516pt,bbury=764pt,height=82mm,width=83mm,angle=-90,clip=}
\end{picture}\par
\vspace{0.2cm}
\caption{\label{ellpa1}
B surface brightness ($mag/arcsec^2$) for population I (stars) and  II 
(crosses) and for the whole galaxy (ellipses) as a function of the 
semi-major axis length, in arcsec.  
The dashed line shows the level of the night sky brightness in B.}
\end{minipage}
\end{figure}

We now apply the above method 
to the central region of NGC 3077 ($\sim$1 kpc) with the aim to get the 
distribution of the young population I and its age distribution and extinction.
We use  intrinsic colours for the young population from the population 
synthesis model data of Bressan et al. 1994, 
 choosing a metallicity value of z=0.05,  
to put a formal metallicity difference between the two populations. 
(For the old population II solar metallicity, z=0.02, was assumed throughout 
this paper following Chromey 1974). 
The influence of this choice for the metallicity on the results is minor, however, since 
the difference of the age lines in the region of interest -- the young 
populations -- is small, as can be seen from Figures \ref{mt2} and  \ref{syP}. 


First we analyse  the  overall contribution of  both 
populations I and II to the observed B light  using the  intensity ratio
($\beta$)  and its complementary, (1-$\beta$).
 These distributions were fitted by isophote ellipses using a Fortran program to extract the characteristic parameters of the two populations.
 Figure \ref{ellpa1} gives the surface brightness distribution of both populations I and II in comparison with that of the whole galaxy.
The contribution of  population I is highest in the centre,
 steeply decreasing  with distance from the centre. 
Population II is present also 
in the central part but becomes the dominant population in the outer part.

\begin{figure}
\begin{minipage}{115mm}
\psfig{figure=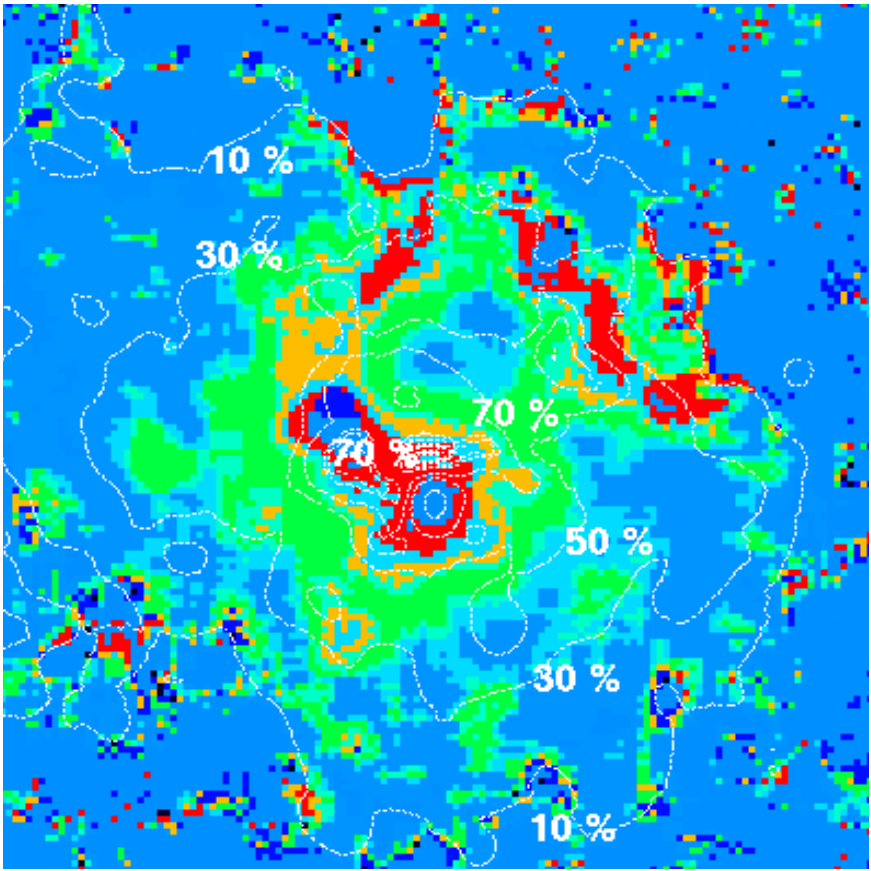,bbllx=144pt,bblly=366pt,bburx=496pt,bbury=732pt,height=115mm,width=115mm,angle=0,clip=}
\end{minipage}
\hfill
\parbox[t]{55mm}{
\caption{The distribution of the  age of population I in the central region of 
NGC 3077 (crudely quantizised). The darkest spot in the middle left side 
is a prominent star cluster with age of 4 Myr.
Age coding is as follows: dark grey (inner regions)  (13 Myr), light  grey
(inner regions) 
(25 Myr), middle grey (50 Myr), light grey (outer regions) (80 Myr) and 
dark grey outer  background (100 Myr).
 Contours superposed on the age distribution show the contribution of 
population I to  the observed B light. North is up and east is 
to the left. The image size is 1.5x1.5 kpc.}
\label{beta}}
\end{figure}

 Figure \ref{beta} shows the age distribution of the population I, as obtained
 using the mentioned method. The Figure depicts an interrupted  ring structure 
of  a very young population,  age between  13 and  50 Myr (follow the inner 
dark grey structure in the figure). 
 A spot of age 4 Myr is embedded in the left part of the ring (the darkest 
spot in the figures). This spot lies at the position of the prominent young star
cluster in NGC 3077 mentioned by  Price and Gullixson 1989. It can also
 immediately be seen on  an  HST image.

This structure is superposed on a homogeneous background distribution of 
an older population I aged about 
 100 Myr (the outer dark  grey  background). This distribution suggests that 
several successive  recent star formation events  occured  in the central 
region of NGC 3077.  

 The contribution of the population I in the observed light shown by the 
dashed contours in  Figure \ref{beta} is highest  
in the centre and becomes lower going
 outward to reach 10\% at nearly $60^{\prime\prime}$ from the centre
 (the boundary of the central region of NGC 3077, see also HN1).  

\begin{figure}
\begin{minipage}{115mm}
\psfig{figure=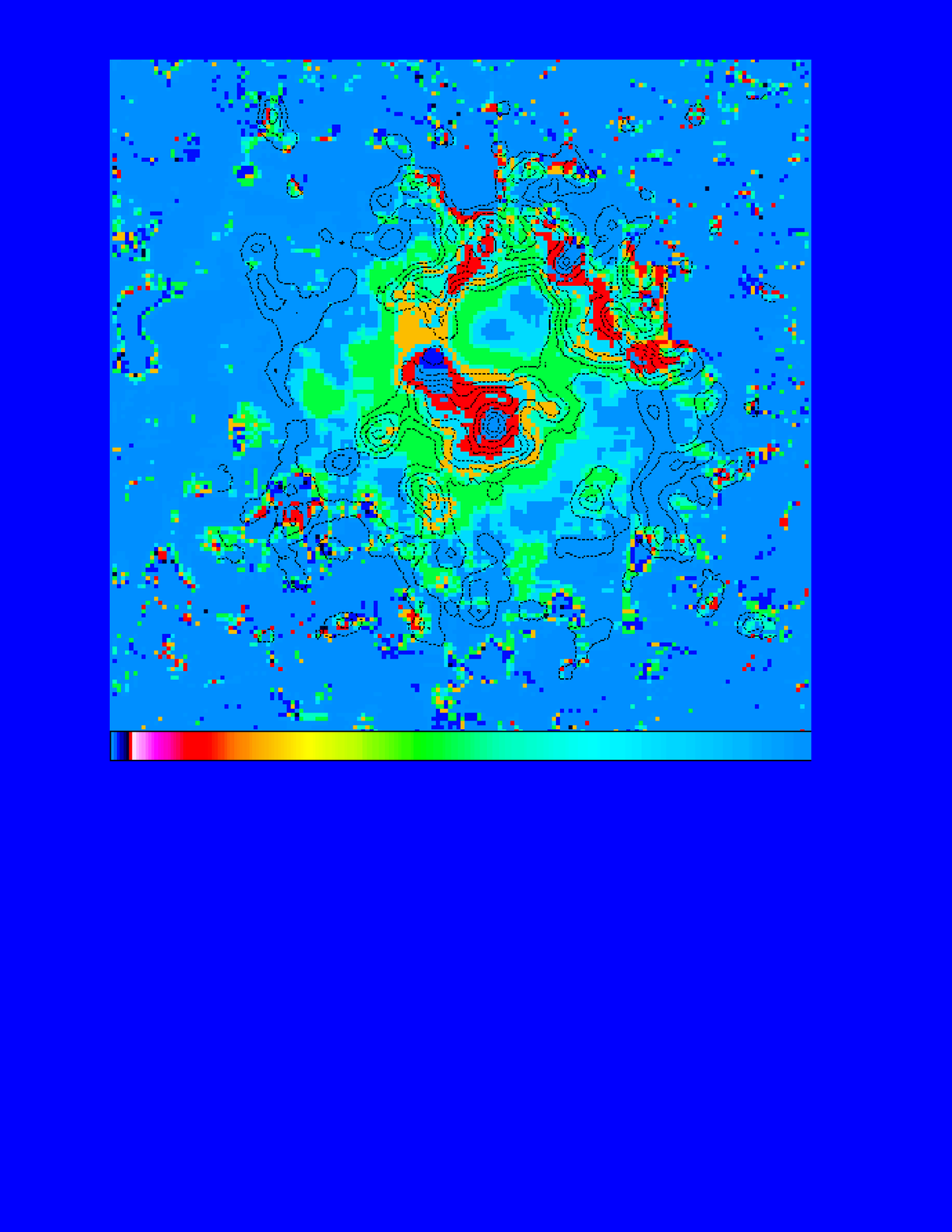,bbllx=144pt,bblly=366pt,bburx=496pt,bbury=732pt,height=115mm,width=115mm,angle=0,clip=}
\end{minipage}
\hfill
\parbox[t]{55mm}{
\caption{The $E_{b-v}$ contours from 0.1 to 0.5, step 0.1 superposed on the age distribution of population I in the central region. North is up  and east is to the left. The image size is 1.5x1.5 kpc. }
\label{ebv}}
\end{figure}

The  extinction distribution of population I in the central region of NGC 3077
is shown in Figure \ref{ebv} in terms of $E_{b-v}$. 
$E_{b-v}$ isophotes from 0.1 to 0.5 step 0.1 mag are superimposed on the 
age variation image.  They are correlated with the ring structure of the age distribution.
A high extinction exists at south-west of the young star-cluster 
(the dark spot). At this location lies the prominent dark cloud in NGC 3077.


\section{Conclusion}
 We  present a simple method for deducing the relative contributions of the 
stellar populations of galaxies by an analysis of their colours.  
We solve the 'mixing equations' (by iteration) for the case of 
two populations of markedly different age in NGC 3077.

For NGC 3077, the main results of this paper can be summarized as follows:\\
- We derive the spatial distribution
of both populations together with an estimate of  their  age and an
 extinction  estimate for population I. \\
- The young population is concentrated in the central region of NGC 3077.
Its contribution to  the observed light decreases 
going out from the centre.\\
- The youngest structures, aged 10-20 Myr, concentrate along a  ringlike 
structure,  associated with dust. 
The ring breaks up into discrete star-forming clumps.\\
- An age gradient for the young population in the central region of NGC 3077 is
found,  ages range from 4 Myr in the centre to 100 Myr at a projected radial 
distance of 1 kpc. Such age variation reflects the occurence of  different 
events of star formation, probably starting from the beginning of the 
encounter with M81 (a few hundred million years) until now.\\
- The prominentest star cluster in the central region is found to have the 
lowest age, about 4 Myr. 

The abovementioned results encourage us to use our method to study the stellar populations in faint galaxies using only the familiar broad band colours.

\acknowledgements

This work is  part of the  dissertation of H.A., financially supported by a
DAAD grant, and carried out  in the Astrophysikalisches Institut Potsdam (AIP).
H.A. gratefully appreciates the kind and fruitful hostpitality in AIP.
This research has made use of NASA/IPAC Extragalactic Database (NED), which
is operated by the Jet Propulsion Laboratory, Caltech, under contract 
with NASA.

\refer
\aba
\rf{Abel-Hamid H.A. and  Notni P.: 2000, Astron. Nachr., this issue. (HN1)}
\rf{Abdel-Hamid H.A.: 1998, Dissertation, Uni. Potsdam, Wissensch.-Verlag Berlin.}
\rf{Abraham, R.G., Ellis, R.S., Fabian, A.C., Tanvir, N.R., and Glazebrook, K.: 1999, MNRAS 303, 641.} 
\rf{Almoznino E. and Brosch N. :1998, MNRAS 298, 931.}
\rf{Bell, E.F., Bower, R.G., deJong, R.S.,  Hereld, M. and Rauscher, B.J.:1999, MNRAS 302, L55}
\rf{Bressan A., Chiosi C. and Fogotto F.: 1994, ApJSupp 94, 63. (BCF94)}
\rf{Chromey F. R.: 1974, MNRAS 174, 455.}
\rf{Grebel E.K. and  Roberts W.J.: 1995, ApJSupp 109, 293.}
\rf{Kong, Xu et al. (28 authors): 2000, Astron J. 119, 2745}
\rf{Larson, R.B. and Tinsley, B.M.: 1978, Astrophys. J. 219, 46. (LT78)}
\rf{Notni P. and  Bronkalla W.: 1984, Astron. Nachr. 305(4) 157.}
\rf{Price J. S. and  Gullixon C.A.: 1989, Astrophys. J. 337, 658.}
\abe

%
\addresses
Hamed Abdel-Hamid\\
National Research Institute of\\
Astronomy and Geophysics\\
11421 Helwan, Cairo\\
Egypt.\\
E-mail hamid@nriag.sci.eg {\em or\/} hamed\_a2000@yahoo.com\\
\\
P. Notni\\
Astrophysikalisches Institut Potsdam\\
An der Sternwarte 16\\
D-14482 Potsdam\\
Germany\\
E-mail pnotni@aip.de \\
\end{document}